\documentclass[12]{article}

\begin{document}

\author{Preston Jones \\
Department of Mathematics and Physics\\
University of Louisiana at Monroe\\
Monroe, LA 71209}
\title{Electromagnetic radiation from temporal variations in space-time and
progenitors of gamma ray burst and millisecond pulsars}
\date{August 16, 2007}
\maketitle

\begin{abstract}
A time varying space-time metric is shown to be a source of electromagnetic
radiation even in the absence of charge sources. The post-Newtonian
approximation is used as a realistic model of the connection between the
space-time metric and a time varying gravitational potential. Rapid temporal
variations in the metric from the coalescence of relativistic stars are
shown to be likely progenitors of gamma ray burst and millisecond pulsars.
\end{abstract}

\section{Introduction}

With the observation of the bending of light around the sun by Eddington in
1919 it has been generally accepted that gravity and electromagnetism are
interrelated. In the intervening years this relation between gravity and
electromagnetism has been established beyond the association of gravity with
curvature of space-time and the path of a beam of light. For example recent
research \cite{Moortgat} has established a theoretical connection between
the gravity wave predictions of general relativity and stimulation of
electromagnetic radiation in charged plasmas. A more fundamental
relationship between gravity and electromagnetic radiation will be
established here by solving the covariant Maxwell's equations in the
presence of a rapidly time varying gravitational potential and in the
absence of any charge sources. In particular rapid variations in space-time,
as a consequence of the collision of astronomical objects, will be shown to
be a source of observed electromagnetic emissions.

It is well established that the homogeneous and inhomogeneous 4-space
equations of electrodynamics reduce to the familiar four Maxwell equations
for the special case of a Lorentz inertial frame and flat space-time metric.
Where the space-time is not flat these equations are complicated by
space-time variations in the metric. This complication follows from the
appearance of the metric in the relation between the electric and magnetic
fields and the differential 4-space operator in the 4-space equations.
Because of this explicit dependence on the metric in the 4-space equations
of electrodynamics variations in space-time would be expected to contribute
to the electrodynamic fields. It would also be expected that temporal
variations in the metric will contribute to the electric and magnetic fields
independent of the local charge distribution.

The 4-space equations of electrodynamics, in order to insure the invariance
of the equations, must relate the 4-space invariant electric and magnetic
fields and differential operator to the invariant charge sources. This is
most commonly achieved by first relating a 2-rank tensor, the
Electromagnetic Tensor, to the electric and magnetic fields \cite{Jackson}.
The association between the Electromagnetic Tensor, the 4-space metric, and
the electromagnetic fields in the equations of electrodynamics is then
somewhat obscure since the fields only appear indirectly in the equations.
It is much easier to identify the relation between the electric and magnetic
fields and the 4-space metric in an alternative representation of the
equations of electrodynamics that includes the fields directly in the
invariant equations. The explicit appearance of the fields in the 4-space
equations was established by Ellis \cite{Ellis}\cite{Jones} where the
electromagnetic tensor is replaced with the direct products of the fields
and the local 4-velocity.

In order to examine the relation between a time varying space-time metric
and the electromagnetic fields the 4-space electrodynamic equations are
expanded here under the conditions of a rapidly time varying metric in a
space-time region that is free of charge sources. Solving for the divergence
of the electromagnetic power the time variations in space-time are shown to
be a source of electromagnetic radiation. These rapid time variations in the
space-time metric can be related to realistic astronomical phenomena by
expressing the metric in terms of the gravitational potential using the
post-Newtonian approximation. From this approximation the rapid changes in
the gravitational potential, associated with collisions of black-holes, are
shown to be a probable source for the power spectra from gamma ray burst. As
a second example of this phenomenon, temporal changes in the gravitational
potential during ring-down following neutron star collisions are shown to be
the possible progenitor of millisecond pulsars.

\section{Covariant Equations of Electrodynamics}

The equations of electrodynamics developed by Ellis \cite{Ellis} exhibit an
implicit connection between the electric and magnetic fields and the 4-space
metric through the covariant derivatives. Following Ellis \cite{Ellis} \cite
{Sonego98}, the homogeneous Maxwell equations can be expressed in covariant
form as the direct product of the fields and the local 4-velocity, 
\begin{equation}
\left[ B^{\gamma }u^{\delta }\right] _{;\gamma }-\left[ u^{\gamma }B^{\delta
}\right] _{;\gamma }+\left[ e^{\gamma \delta \alpha \beta }u_{\alpha
}E_{\beta }\right] _{;\gamma }=0.  \label{1st Maxwell}
\end{equation}

\noindent The 4-velocity $u_{\alpha }$ is the velocity of the differential
volume element associated with the fields and $B^{\gamma }$ and $E_{\beta }$
are the components of the 4-vector magnetic and electric fields
respectively. The Levi-Civita tensor $e^{\alpha \beta \gamma \delta }$ is
defined in terms of the permutation symbol $E^{\alpha \beta \gamma \delta }$%
, which is zero if any indices are repeated, one for even permutations of 0,
1, 2, 3, and negative one for odd permutations,

\begin{equation}
e^{\alpha \beta \gamma \delta }=-\frac{1}{\sqrt{-g}}E^{\alpha \beta \gamma
\delta },  \label{Levi-Civita}
\end{equation}
where $g=\det \left( \mathbf{\hat{e}}_{\alpha }\cdot \mathbf{\hat{e}}_{\beta
}\right) $ is the determinant of the metric components. The inhomogeneous
Maxwell equations are written in a similar fashion \cite{Sonego98},

\begin{equation}
\left[ E^{\gamma }u^{\delta }\right] _{;\gamma }-\left[ u^{\gamma }E^{\delta
}\right] _{;\gamma }-\left[ e^{\gamma \delta \alpha \beta }u_{\alpha
}B_{\beta }\right] _{;\gamma }=4\pi J^{\delta },  \label{2nd Maxwell}
\end{equation}

\noindent where $J^{\delta }$ is the 4-vector electric current density.
Assuming a Lorentz inertial frame (LIF) the covariant derivatives can be
replaced with partial derivatives. Adopting the convention of a negative
time part for the metric the 4-velocity is $u_{\alpha }=u_{0}=-1$. Expanding
the homogeneous equation with $\delta =0$ leads to 
\begin{equation}
B_{,k}^{k}=0
\end{equation}

\noindent and in the inhomogeneous equation to

\begin{equation}
E_{,k}^{k}=4\pi J^{0},
\end{equation}

\noindent where the Roman indices range over $1,\,2,\,3$. Taking $\delta =i$
in the homogeneous and inhomogeneous equations produces the remaining
Maxwell equations, 
\begin{equation}
B_{,0}^{i}+E^{ijk}E_{k,j}=0
\end{equation}

\noindent and

\begin{equation}
E^{ijk}B_{k,j}-E_{,0}^{i}=4\pi J^{i}.
\end{equation}

\noindent This expansion demonstrates that the covariant equations \ref{1st
Maxwell} and \ref{2nd Maxwell} are the correct form invariant equations of
electrodynamics.

\section{Time Dependent Field in Post Newtonian Approximation}

In order to examine the relation between the covariant equations of
electrodynamics and variations in the gravitational potential the 4-space
metric will be expanded in terms of the potential. The Post-Newtonian
approximation assumes that only the lowest order terms in this expansion
contribute appreciably to the metric. Making this assumption and requiring
that the gravitational potential be time dependent, the time component of
the metric is

\begin{equation}
g_{00}=-\left( 1+2\phi \right) ,\quad g^{00}=-\left( 1-2\phi \right) .
\label{metric time}
\end{equation}

\noindent In this post-Newtonian approximation the spacial components are,

\begin{equation}
g_{ii}=1-2\phi ,\quad g^{ii}=1+2\phi ,  \label{metric space}
\end{equation}

\noindent with all other components equal to zero. The Christoffel symbols
will be needed in the expansion of the covariant derivatives and the nonzero
terms are

\begin{equation}
\Gamma _{00}^{0}=-\Gamma _{ii}^{0}=\dot{\phi}-2\phi \dot{\phi},\quad \Gamma
_{i0}^{i}=-\dot{\phi}-2\phi \dot{\phi}.
\end{equation}

\noindent where the dots represent time derivatives and there is no sum on $%
i $. It will be helpful later to define a sum of two Christoffel symbols,

\begin{equation}
\Gamma _{00}^{0}+\Gamma _{i0}^{i}=-2\frac{\partial }{\partial t}\phi
^{2}=-\Sigma .
\end{equation}

\noindent In a comoving frame, $u_{i}=0$, the covariant derivatives in the
equations of electrodynamics can be expanded in terms of this sum. The time
derivatives in terms of some arbitrary vector $A^{k}$ and the direct product
with the 4-velocity are

\begin{equation}
\left[ u^{0}A^{k}\right] _{;0}=A_{,0}^{k}-\Sigma A^{k}.  \label{time diff}
\end{equation}

\noindent The covariant space derivatives are simply the partial derivatives,

\begin{equation}
\left[ u^{0}A^{k}\right] _{;k}=A_{,k}^{k}.  \label{space diff}
\end{equation}

\noindent The dual term in the equations requires a bit more attention.
Since the metric is not position dependent the Christoffel symbols will
vanish and the covariant derivatives are again partial derivatives,

\begin{equation}
\left[ e^{ij0k}u_{0}A_{k}\right] _{;i}=\left[ \left( -\frac{1}{\sqrt{-g}}
E^{ij0k}\right) \left( -1\right) A_{k}\right] _{,i}.
\end{equation}

\noindent With our assumptions of a time dependent metric in the
post-Newtonian approximation the spacial derivatives of the determinate make
no contribution and to first order the covariant derivatives of the dual
terms,

\begin{equation}
\left[ e^{ij0k}u_{0}A_{k}\right] _{;i}=-E^{ijk}A_{k,j},
\end{equation}

\noindent are simply the negative of the curl.

\section{Electromagnetic Power Spectra}

The post-Newtonian approximation for a time varying 4-space metric will be
used to demonstrate the connection between the gravitational potential and
the electromagnetic fields in the covariant equations of electrodynamics.
The equations of electrodynamics will be expanded assuming a comoving
reference frame and small second order terms in the metric, $\phi ^{2}\prec
\prec 1$ and $\phi ^{2}\prec \prec \frac{\partial }{\partial t}\phi ^{2}.$
To demonstrate that a time varying gravitational potential generates
electromagnetic radiation assume also that there are no charge sources, $%
J^{\alpha }=0$. Taking the equations for $\delta =i$ and no charge sources, 
\begin{equation}
\left[ u^{0}B^{i}\right] _{;0}+E^{ijk}E_{k,j}=0
\end{equation}

\noindent and

\begin{equation}
\left[ u^{0}E^{i}\right] _{;0}-E^{ijk}B_{k,j}=0.
\end{equation}

\noindent Making use of the expansions of the covariant derivatives from the
previous section and contracting each equation with the appropriate field,

\begin{equation}
B_{i}B_{,0}^{i}+B_{i}E^{ijk}E_{k,j}=\Sigma B^{2}
\end{equation}

\noindent and 
\begin{equation}
E_{i}E_{,0}^{i}-E_{i}E^{ijk}B_{k,j}=\Sigma E^{2}.
\end{equation}

\noindent Adding these two equations and rearranging the terms, 
\begin{equation}
B_{i}E^{ijk}E_{k,j}-E_{i}E^{ijk}B_{k,j}+\frac{1}{2}\frac{\partial }{\partial
t}\left( B^{2}+E^{2}\right) =\Sigma \left( B^{2}+E^{2}\right) .
\end{equation}

\noindent The terms on the left hand side can be simplified using the triple
product identity, $\nabla \cdot \left( \mathbf{a\times b}\right) =\mathbf{b}%
\cdot \left( \nabla \mathbf{\times a}\right) -\mathbf{a}\cdot \left( \nabla 
\mathbf{\times b}\right) $, dividing by $\frac{1}{4\pi },$ and substituting
the energy density $U=\frac{1}{8\pi }\left( E^{2}+B^{2}\right) $,

\begin{equation}
\nabla \cdot \mathbf{S}+\frac{\partial U}{\partial t}=\frac{4}{c^{4}}U\frac{%
\partial }{\partial t}\phi ^{2},  \label{Poynting}
\end{equation}

\noindent where use has been made of the Poynting vector, $\mathbf{S=}\frac{c%
}{4\pi }\mathbf{E\times B,}$ and the factor of $c$ has been reinstated to
illustrate the order of magnitude of this effect. This demonstrates that in
a region of nonzero internal energy density $U$ and rapidly time varying
gravitational potential, $\phi ,$ electromagnetic radiation will be
generated. Due to the square of the gravitational potential in the time
derivative the period of the electromagnetic power spectra would be half the
period of the variation in the gravitational potential. This relationship
between electromagnetic radiation and a time varying metric is formally
similar to the expression for the interaction between gravity waves and
charged plasmas \cite{Moortgat}. The current development is a more general
consequence of the covariant derivatives in the Maxwell equations and in
particular does not assume a charged plasma or any other charge sources.

\section{Gamma Ray Burst and Millisecond Pulsars}

Let us consider the electromagnetic radiation that will be generated from
the collision of relativistic stars. Relativistically the
internal energy density $U$\ would be associated with the total mass-energy
density of the binary system. During coalescence a fraction of the rest mass
energy will be radiated from the system as electromagnetic radiation. Since
the gravitational potential increases most rapidly at the end of the
collision process this is a likely progenitor of the radiation spike seen in
gamma ray burst. Collisions of neutron star binaries have been previously
identified as a potential source of gamma ray burst \cite{Moortgat}\cite
{Meszaros}\cite{Allen99}. However, the magnitude of the radiated power
calculated from the collisions of neutron stars is much smaller than the
observed radiated power.

Since the typical energy radiated from gamma ray burst is much greater than
the internal energy density of neutron star binaries the progenitor of the
these burst are much more likely to be black hole collisions. Hawking \cite
{Hawking} has estimated that as much as 50 percent of the original rest mass
energy escapes during the coalescence of black hole binaries. Hawking's
estimate would place a lower bound on the radiated energy of the order of
one solar mass. However, the internal energy associated with a black hole
binary does not have an upper bound and explains the magnitude of the energy
radiated in even the largest gamma ray burst. The period of the coalescence
of black hole binaries should be associated with the interval of the
transition from two separate event horizons to a single event horizon which
is consistent with the short duration of the peak energy of gamma ray burst.
This association between black holes and gamma ray burst provides a method
of determining the order of magnitude of the masses of black holes and could
also provide a method for determining the distribution of black holes in the
universe and the associated mass distribution.

Gamma ray bursts are one example of an electromagnetic power spectra that
would be associated with an astronomical event involving a rapidly varying
gravitational potential. However, the generation of electromagnetic
radiation by time variations in a gravitational potential should be
ubiquitous. In the example of the collision of a neutron star binary there
is also the possibility of a ring-down of the coalesced object, provided
that the combined mass does not produce a black hole. During this ring-down
the gravitational potential would be expected to generate an electromagnetic
power spectra due to the time variation in the square gravitational
potential. The pulsation modes of the metric and associated gravitational
potential, in the ring down of a coalesced neutron star, have been
calculated by Allen et al \cite{Allen99}\cite{Allen98}. The calculated modes
predict millisecond time variations in the gravitational potential which
would explain the mechanism for electromagnetic radiation from millisecond
pulsars.

\section{Conclusion}

Expanding the covariant equations of electrodynamics, in terms of the direct
products of the electromagnetic fields and the local 4-velocity and assuming
the post-Newtonian approximation, a time varying metric was shown to be a
source of electromagnetic radiation even in the absence of charge sources.
Electromagnetic radiation produced by rapid time variations in the
gravitational potential of coalescing black holes is the likely progenitor
of gamma ray burst. This gravity induced electromagnetic radiation is also
the likely mechanism for producing the radiation from millisecond pulsars.
The phenomena of gravity induced electromagnetic radiation would not be
expected to be limited to these two examples. However, due to the factor of $%
c^{4}$ in the relation between the radiated power and the time rate of
change of the potential this phenomenon will be observable only in
astronomical events where space-time variation in the potential and the
internal energy are great.

\end{document}